\documentclass[twocolumn]{article}
\usepackage{amsmath,amssymb}
\usepackage{graphicx}
\usepackage{authblk}
\usepackage{slashed}
\usepackage{cite}
\usepackage{hyperref}
\usepackage{flushend}
\usepackage{xfrac}

\begin{document}
\title{
%\vspace{-3.0cm} 
%{\normalsize  DCPT/NUMBER; IPPP/NUMBER \hfill\mbox{}\hfill\mbox{}}\\
\vspace{-1cm} 
\large{\textbf{
Making The Most Of MET: Mass Reconstruction From Collimated Decays
}}}

\author[1]{Michael Spannowsky}
\author[1,2]{Chris Wymant}
\affil[1]{\small{\it{Institute for Particle Physics Phenomenology, Department of Physics\\
Durham University, Durham DH1 3LE, United Kingdom}}}
\affil[2]{\small{\it{Laboratoire de Physique Th\'{e}orique\thanks{}, Universit\'{e} de Paris-Sud XI, B\^{a}timent
210, F-91405 ORSAY Cedex, France
}
}}
\date{}
\twocolumn[
  \begin{@twocolumnfalse}
    \maketitle
\vspace*{-10mm}
    \begin{abstract}
At hadron colliders invisible particles $\chi$ can be inferred only through observation of the transverse component of the vectorial sum of their momenta -- missing $E_T$ or MET -- preventing reconstruction of the masses of their mother particles.
Here we outline situations where prior prejudice about the event kinematics allows one to make the most of MET by decomposing it into its expected sum of transverse contributions, each of which may be promoted to a full four-momentum approximating the associated $\chi$.
Such prejudice arises when all $\chi$ in the event are expected to be light and \mbox{(anti-)parallel} to a {\it visible} object, due to spin correlations, back-to-back decays or boosted decays.
We focus on the last of these, with boosted semi-invisibly decaying neutralinos widely motivated in supersymmetry (in the presence of light gravitinos, singlinos, photini or pseudo-goldstini), and demonstrate our simple method's ability to reconstruct sharp mass peaks from the MET decomposition.
    \end{abstract}
  \end{@twocolumnfalse}
  \vspace*{8mm}
  ]

{
  \renewcommand{\thefootnote}%
    {\fnsymbol{footnote}}
  \footnotetext[1]{Unit\'{e} Mixte de Recherche (CNRS) UMR 8627}
}

%%%%%%%%%%%%%%%%%%%%%%%%%%%%%%%%%%%%%%%%%%%%%%%%%%%%%%%

\section{Introduction} \label{setup}
Missing transverse energy or MET is of great importance at hadron colliders: it is our only way of inferring the presence of non-interacting (collider-)stable particles $\chi$.
However whenever {\it two} such particles are produced (which will always be the case if their stability is due to a $\mathbb{Z}_2$ symmetry) our observation only of the vectorial sum of their transverse momenta $\slashed{E}_T = |\slashed{\mathbf{p}}_T| = |\mathbf{p}_{a,T} + \mathbf{p}_{b,T}|$ thwarts reconstruction of masses in the decay cascade\footnote{
$2\chi$ could also be directly produced, giving a final state with, at leading order, no large transverse energy (visible or invisible).
The universal possibility of hard initial state radiation allows essentially model-independent limits to be set on the direct production of new $\chi$ particles from monojet and monophoton searches.
Here we focus only on production of $2\chi$ via a decay cascade.
}
ending with $2\chi$.
Popular methods for searching for heavy particles with partially invisible decays are transverse mass observables \cite{Barger:1987re}, $M_{T2}$ \cite{Lester:1999tx}, razor analyses\cite{Rogan:2010kb} and kinematic edges \cite{Allanach:2000kt}.

We start by asking under what circumstances can we explicitly access the missing momentum associated with each $\chi$ particle separately.
Clearly some feature of the rest of the event must suggest the correct decomposition of $\slashed{\mathbf{p}}_T$ into $\mathbf{p}_{a,T} + \mathbf{p}_{b,T}$.
If there are two well-localised visible objects that we expect, from some prior prejudice about the kinematics, to be parallel or anti-parallel to the two unseen $\chi$ particles, then we have two directions in the transverse plane to give us $\mathbf{p}_{a,T}$ and $\mathbf{p}_{b,T}$.
Furthermore we can add longitudinal components to each of these two transverse vectors to make them \mbox{(anti-)parallel} to their corresponding visible object in three dimensions, giving approximations for $\mathbf{p}_{\chi_{a,b}}$.
If $\chi$ is much lighter than the particle produced in the hard scattering, i.e. at the start of the decay cascade, we can promote $\mathbf{p}_{\chi_{a,b}}$ to massless four-vectors; we will show that combined with the four-vectors for the visible decay products, a strong mass peak for the initial particles can be reconstructed.

\section{Motivation}

Parallel or anti-parallel visible and missing energy is not worth considering only for its ease: it can arise in many circumstances.
Spin correlations may make $\chi$ particles approximately \mbox{(anti-)parallel} to other particles.
Two-body decays of particles $P$ not boosted in the lab frame, such as the majority of those produced directly, are usually back-to-back: therefore in $2P\rightarrow2\chi+2\text{vis}$, each $\chi$ is often nearly anti-parallel to one of the `$\text{vis}$'.
(However this simple example lends itself well to numerical optimisation over all possible momenta for the $2\chi$, i.e. to use of $M_{T2}$.)

Of particular interest is when each $\chi$ is produced together with visible energy from the decay of a {\it boosted} particle.
This will arise whenever a) directly pair-produced particles are appreciably heavier than whatever they decay into in the first step of the cascade, and b) $\chi$ are created following two or more steps.
Together these points imply that each of the two `sides' of the event (separated according to the mother particle) contains an intermediate particle which is boosted: the visible object(s) and $\chi$ it ultimately decays to will be collimated.

For some examples, consider the quintessential supersymmetry (susy) decay of a pair-produced squark to a hard jet and {\it light} neutralino : $\tilde{q}\rightarrow q+\tilde{N}_1$ (we denote the susy neutralinos by $\tilde{N}_i$ to avoid confusion with our generic invisible particle $\chi$).
There are many reasons why we might expect $\tilde{N}_1$ to be unstable, decaying to visible energy and a lighter, neutral, collider-stable particle -- the latter could be:
\begin{itemize}
\item a gravitino $\tilde{G}$, if susy breaking is mediated at a low scale, i.e. some form of gauge mediation.
A low mediation scale is motivated by electroweak naturalness and an automatic solution of the susy flavour problem.
See~\cite{Giudice:1998bp} for a review and~\cite{Kats:2011qh} for a comprehensive list of possible collider signatures.
\item a pseudo-goldstino $\tilde{G'}$, if more than one hidden sector breaks susy, as may occur in string theory or quiver gauge theories
\cite{Cheung:2010mc,Argurio:2011hs}.
See~\cite{Argurio:2011gu} for the collider phenomenology.
\item a singlino\footnote{
In this case the decay is not really $\tilde{q}\rightarrow q+\tilde{N}_1\rightarrow\ldots$ but $\tilde{q}\rightarrow q+\tilde{N}_2\rightarrow\tilde{N}_1+\ldots$, since new photini/singlinos actually mix with the MSSM neutralinos. If $\tilde{N}_2$ is mostly `MSSM-like' (any mixture of Higgsino, wino and bino), and $\tilde{N}_1$ is mostly singlino or a new photino, then direct decay of $\tilde{q}$ to $\tilde{N}_1$ is suppressed relative to the two-step decay.
 } $\tilde{S}$, if the Minimal Supersymmetric Standard Model (MSSM) is extended with a gauge-singlet superfield, giving the Next-to-MSSM (NMSSM).
This is motivated by the $\mu$-problem of the MSSM, and also by naturalness~\cite{Hall:2011aa}.
See~\cite{Ellwanger:2009dp,Maniatis:2009re} for a review and~\cite{Das:2012rr} for the modified collider signals.
\item a new photino\footnotemark[\value{footnote}] $\tilde{\gamma}'$, if the MSSM is extended with one or more extra U(1) gauge symmetries, as is commonly expected to arise from string compactifications.
See~\cite{Baryakhtar:2012rz} for a discussion of collider prospects.
\end{itemize}
In the nomenclature of \cite{Baryakhtar:2012rz}, $\tilde{N}_1$ here is the Lightest {\it Ordinary} Supersymmetric Particle (LOSP).
All of these examples have some other particle as the true LSP, and so a charged or coloured susy particle could be lighter than $\tilde{N_1}$ and take its place as the LOSP in the cascade $\tilde{q}\rightarrow \text{vis}_1+(\text{LOSP})\rightarrow\text{vis}_1+(\text{vis}_2+\text{LSP})$, giving different visible energy.

\section{The Analysis} \label{sec:Analysis}
We elaborate on the strategy outlined in Section \ref{setup} in terms of a concrete example to allow clearer references to the particles involved in the signal: we consider the classic gauge-mediation decay\footnote{
A similar final state may arise from Universal Extra Dimensions \cite{Macesanu:2002db}, though semi-invisibly decaying Kaluza-Klein photons from KK quark/gluon decays are not generally expected to be boosted; this will be important for our analysis.}
$2\tilde{q}\rightarrow 2q+2(\tilde{N}_1)\rightarrow 2q+2(\tilde{G}+\gamma)$.
The lightest neutralino is typically expected to be considerably lighter than the squarks in this scenario, as the phenomenon of gaugino screening in the simplest models makes the gauginos much lighter than the scalars (see e.g.~\cite{Cohen:2011aa}) and renormalisation-group evolution tends to drive squark masses up and the bino mass down.
This simple observation gives a powerful handle on the signal, as yet unexploited: the gravitinos and photons are normally collimated.
It is exploited as follows.
\begin{enumerate} \itemsep0pt \parskip0pt \parsep0pt
\vspace*{-1mm}
{\setlength\itemindent{-2pt}  \item Uniquely decompose $\slashed{\mathbf{p}}_T$ into $\mathbf{p}_{a,T} + \mathbf{p}_{b,T}$ which are defined to be parallel, in the transverse plane, to the two hardest isolated photons.
}
{\setlength\itemindent{-2pt}  \item Promote $\mathbf{p}_{a,b;T}$ to three-vectors $\mathbf{p}_{a,b}$ by adding the longitudinal components required to make them parallel to each of the photons in three dimensions.
}
{\setlength\itemindent{-2pt}  \item Promote $\mathbf{p}_{a,b}$ to massless four-vectors $p_{a,b}^\mu = (|\mathbf{p}_{a,b}|, \mathbf{p}_{a,b})$, giving approximations for the two gravitino four-vectors.
Adding each of these to the four-vector of the collinear photon gives massless approximations for the two neutralino four-vectors, $p_{\tilde{N}_{1;a,b}}^\mu$.
}
{\setlength\itemindent{-2pt}  \item If each neutralino $\tilde{N}_{1;a,b}$ can be paired with the `correct' jet in the event $j_{a,b}$, then taking the invariant mass of each pair reconstructs the mass of the initial squarks: $M_{\text{rec};a,b}^2 =  (p_{\tilde{N}_{1;a,b}}^\mu + p_{j_{a,b}}^\mu)^2$
}
\end{enumerate}

%In \cite{Englert:2012wf} we made use of missing energy decomposition inside a single fat jet from the boosted decay of a Higgs boson to new semi-invisibly decaying (pseudo-)scalars.
Steps 1-2 above reconstruct the three-momenta of the two neutralinos in the same way as is done for the two $\tau$ in $H\rightarrow2\tau\rightarrow e^{\pm}\mu^{\mp}\slashed{E}_T$ with the collinear approximation of~\cite{Plehn:1999xi} (an approximation which we also found to be useful for jet substructure in \cite{Englert:2012wf}).
There, the two $\tau$ four-momenta are added together to get the mass of the single mother particle; here the four-momenta of the two neutralinos  are separately added to those of other visible particles in the event to get the masses of two mother particles -- step 4.

Step 4 needs a criterion for the correct way to pair each reconstructed neutralino with one of the jets in the event, since the mass they should reconstruct is unknown.
The correct jet is considered to be the one most closely resembling the quark produced in the same $\tilde{q}\rightarrow \tilde{N}_1+q$ decay.
Keeping only the two hardest jets, there are two arrangements -- two ways of pairing each neutralino with a different jet.
More generally one can consider the $N$ hardest jets in the event, giving $N(N-1)$ arrangements to choose from.
Each squark is generally produced nearly at rest, therefore the neutralino and jet into which it decays are likely to be back-to-back; the jet is also expected to be hard, with an energy roughly half the squark's mass.
Therefore one criterion is to pair the two neutralinos $\tilde{N}_{1;a,b}$ with jets $j_a$ and $j_b$ so as to make maximally negative the sum of dot products between the three-momenta of each neutralino and its jet:
\begin{equation*}
 \text{criterion } \alpha\!:\; -\!\left( \mathbf{p}_{\tilde{N}_{1,a}}.\mathbf{p}_{j_a} + \mathbf{p}_{\tilde{N}_{1,b}}.\mathbf{p}_{j_b} \right) \; \text{maximal}
\end{equation*}
If the pair-produced squarks are mass degenerate, this can also be exploited: the two reconstructed masses should coincide.
This gives the second possibility for finding the right jets:
\begin{equation*}
 \text{criterion } \beta\!:\; \left| (p_{\tilde{N}_{1,a}}^\mu + p_{j_a}^\mu)^2 - (p_{\tilde{N}_{1,b}}^\mu + p_{j_b}^\mu)^2 \right| \; \text{minimal}
\end{equation*}
Each criterion suggests the correct jets, defining two reconstructed masses $M_{\text{rec};a,b}^2 =  (p_{\tilde{N}_{1;a,b}}^\mu + p_{j_{a,b}}^\mu)^2$.
The maximisation/minimisation above is not differential but discrete -- the quantity is calculated once for each of the $N(N-1)$ arrangements of jets with neutralinos and only the largest/smallest is kept.
It thus takes negligible computational time (indeed $N=2$ is optimal in our example) and could be incorporated into a trigger.
These two criteria are not specific to neutralinos and jets: they are relevant for final states where two objects need to be paired correctly with two other objects, both being the decay products of pair-produced particles (the second criterion also requires mass degeneracy of the two mother particles).

We consider a simplified model with squarks of the first two generations, a bino-like neutralino, and a gravitino with masses $m_{\tilde{q}}= 1.2~\text{TeV},\:m_{\tilde{N}_1}=100~\text{GeV},\:m_{\tilde{G}}=1~\text{eV}$ respectively; this squark mass is at the edge of the strongest current constraints~\cite{CMSsearch}.
We calculate a full spectrum for this simplified model (all other superpartner masses are set $2$~TeV) with
{\tt SoftSusy~3.3.4}~\cite{Allanach:2001kg}
and decay widths with
{\tt Herwig++~2.6.1}~\cite{Bahr:2008pv}.
We then follow two routes to get to observable distributions.
In the first, {\tt MadGraph~5~1.5.5}\cite{Alwall:2011uj} supplies the matrix elements for disquark production; the subsequent decays, extra radiation, showering and hadronisation are done by {\tt PYTHIA~6}~\cite{Sjostrand:2006za}; fast detector simulation is then performed with {\tt PGS~4}~\cite{PGS}.
In the second, {\tt Herwig++} is used to generate the complete event; jets are defined with {\tt FastJet~3.0.3}~\cite{Cacciari:2011ma}, and the final state objects analysed in the {\tt RIVET~1.8.1} framework~\cite{Buckley:2010ar}.
Our kinematical analysis -- steps 1-4 with criteria $\alpha$ and $\beta$ above -- is then applied.
Code for doing this
%, taking as input either {\tt PGS} output or HepMC files,
can be found at~\cite{Me}.

Basic cuts needed for the analysis are as follows:
\begin{itemize} \itemsep0pt \parskip0pt \parsep0pt
\vspace*{-1.5mm}
{\setlength\itemindent{-2pt}  \item At least two jets, clustered using the anti-kt algorithm~\cite{Cacciari:2008gp} with size parameter 0.4.
Jet candidates are required to have $p_T>30$~GeV and $|\eta|<4.5$.
}
{\setlength\itemindent{-2pt}  \item At least two isolated photons with $p_T>10$~GeV.
On the {\tt MadGraph}-{\tt PYTHIA} route, {\tt PGS} handles isolation.
On the {\tt Herwig} route, we consider a photon isolated when the sum of transverse energy in a cone $\Delta R < 0.4$ around the photon is less than $5$~GeV.
}
{\setlength\itemindent{-2pt}  \item A minimum and maximum azimuthal angular separation between the two hardest isolated photons \mbox{$\epsilon < \Delta \phi_{\gamma_1 \gamma_2} < \pi - \epsilon$} with $\epsilon = 0.01$, since photons which are exactly \mbox{(anti-)}parallel in the transverse plane do not allow $\slashed{E}_T$ decomposition.
}
{\setlength\itemindent{-2pt}  \item The missing energy vector $\slashed{\mathbf{p}}_T$ should lie in between the two photons in the transverse plane (i.e. inside the smaller of the two sectors delimited by the two photon directions).
This ensures that the event has $\slashed{\mathbf{p}}_T$ corresponding to the ansatz of both gravitinos being parallel to their photons.
With this cut the kinematics are always in the `trivial zero' of the $M_{T2}$ observable (see ~\cite{Lester:2011nj}).
}
\end{itemize}
Decomposition of $\slashed{E}_T$ of course requires $\slashed{E}_T\neq0$; in practice this is always satisfied.
We do not cut on $\slashed{E}_T$ -- we analyse this particular signal not to optimise the associated cuts but simply as a demonstration of the mass reconstruction technique.
In our present example almost all events have $\slashed{E}_T > 100$~GeV and so a large requirement could be placed as in existing searches (likewise for the leading jet and photon $p_T$ which are typically hard in the signal.)
Note that with a requirement for hard photons and $\slashed{E}_T$ there is typically very small background for new physics~\cite{Kribs:2009yh} and the priority is an observable that increases the visibility of the signal alone, ideally through a resonance.

\begin{center}
\begin{figure}[!ht] 
\centering
\includegraphics[width=0.49\linewidth]{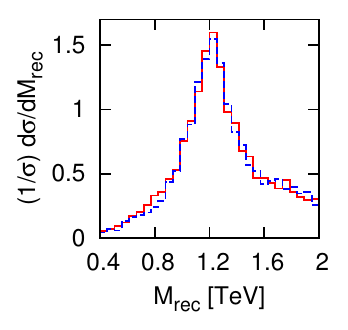}
\includegraphics[width=0.49\linewidth]{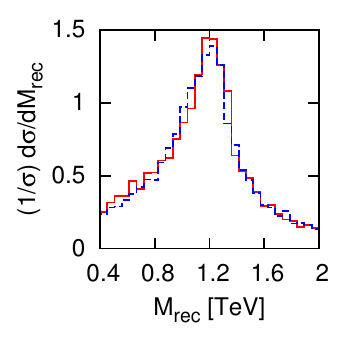}
\includegraphics[width=0.49\linewidth]{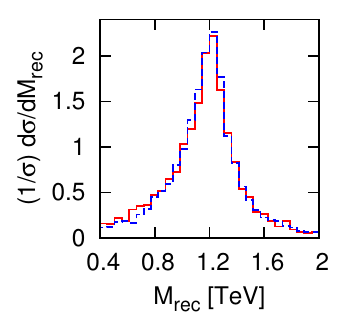}
\includegraphics[width=0.49\linewidth]{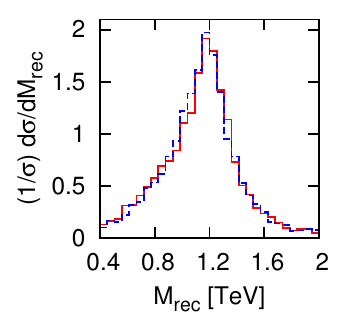}
\caption{
The squark mass ($m_{\tilde{q}}= 1.2~\text{TeV}$) in the process $pp\rightarrow2\tilde{q}\rightarrow 2q+2(\tilde{N}_1)\rightarrow 2q+2(\tilde{G}+\gamma)$ reconstructed with missing energy decomposition and neutralino-jet pairing as described in the text.
The masses of the lightest neutralino and gravitino are $m_{\tilde{N}_1}=100~\text{GeV},\:m_{\tilde{G}}=1~\text{eV}$; the centre of mass energy is $8$~TeV.
Panels on the left (right) show the mass of the squark calculated from the leading (sub-leading) photon in each event.
Upper (lower) panels pair jets with reconstructed neutralinos using criterion $\alpha$ ($\beta$).
The blue dashed line shows events generated by {\tt MadGraph} and {\tt PYTHIA}, with fast detector simulation performed by {\tt PGS}; the red solid line shows events generated by {\tt Herwig++}.
}
\label{fig:1TeV}
\end{figure}
\end{center}

Fig.~\ref{fig:1TeV} shows the results of the analysis.

As our analysis makes use of hard jets arising from the decay of signal particles, it could in principle be affected by the (higher order) production of additional jets in the hard scattering.
To investigate this we simulated $2\tilde{q}$ and $2\tilde{q}+1{\rm jet}$ production and combined these consistently into a single sample using the MLM matching procedure~\cite{Alwall:2007fs}.
The reconstructed mass distributions are essentially identical to those of simple $2\tilde{q}$ production shown in Fig. \ref{fig:1TeV}, which follows from the fact that our method is designed to find the two jets that look most like they have been produced by the decay of the squarks, and other jets are discarded.

\begin{center}
\begin{figure}[!ht]
\centering
\includegraphics[width=0.58\linewidth]{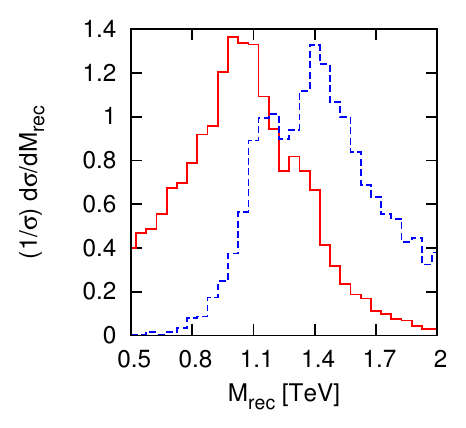}
\caption{
As Fig.~\ref{fig:1TeV} but with one squark from the first two generations having mass $1.1$~TeV and the other seven having mass $1.4$~TeV.
The solid red (dashed blue) line shows the smaller (larger) of the two masses reconstructed in each event.
Only criterion $\alpha$ for jet-neutralino pairing is used.
Events are generated with {\tt Herwig++}.
}
\label{fig:Horns}
\end{figure}
\end{center}

Criterion $\alpha$ can also reconstruct the masses of pair-produced {\it non-degenerate} particles.
In Fig.~\ref{fig:Horns} it is used to analyse the same signal as previously but now with one squark from the first two generations having mass $1.1$~TeV and the other seven having mass $1.4$~TeV.
This unequal splitting is chosen to have large cross-sections for the production of two squarks of {\it different} mass (four lighter squarks and four heavier would merely result in a dominant production of the lighter four alone); nevertheless production of two squarks of the {\it same} mass still has non-zero cross-section.
Thus the distribution of the larger (smaller) of the two masses calculated for each event peaks strongly at $1.4$~TeV ($1.1$~TeV) and weakly at $1.1$~TeV ($1.4$~TeV), with the weak peak resulting from pair-produced degenerate squarks.

\section{Discussion}

\begin{center}
\begin{figure}[!ht]
\centering
\includegraphics[width=0.7\linewidth]{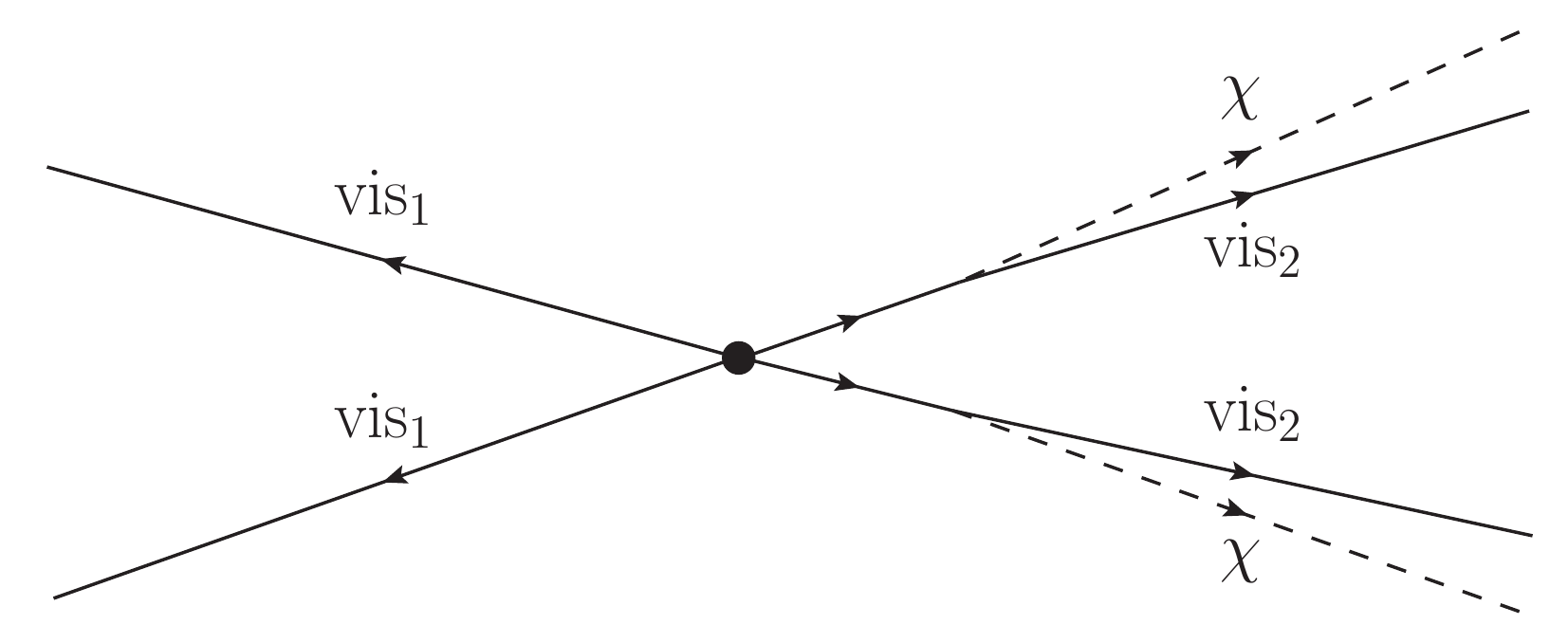}
\caption{
The topology we consider: pair-produced particles each decay into a visible Standard Model particle ${\rm vis}_1$ and a much lighter particle, which is thus boosted; this decays semi-invisibly into ${\rm vis}_2$ and $\chi$.
}
\label{fig:Diagram}
\end{figure}
\end{center}

The final state of the example from the previous section has two jets and two pairs of roughly collinear photons and gravitinos.
The jet could be replaced by any other visible particle -- `vis$_1$' -- the photon too -- `vis$_2$' -- and the gravitino by anything invisible, $\chi$: we show this general topology in Fig.~\ref{fig:Diagram}.  
Provided there are two semi-invisible decays which are boosted (or forced into (anti-)parallel behaviour by spin correlations) the same analysis presented here should in theory have some potential for mass reconstruction.
Of course if vis$_1$ and vis$_2$ are objects less clean experimentally than light-flavour jets and photons, such as $b$ quarks or even combinations of particles, the procedure will be more difficult in practice.
Searches for mass peaks in the manner presented, considering various different particle types for vis$_{1,2}$, could discover expected or unexpected resonances.
Below, we outline how the method might be adapted as the topology is distorted and generalised further.

\begin{center}
\begin{figure}[!ht] 
\centering
\includegraphics[width=0.49\linewidth]{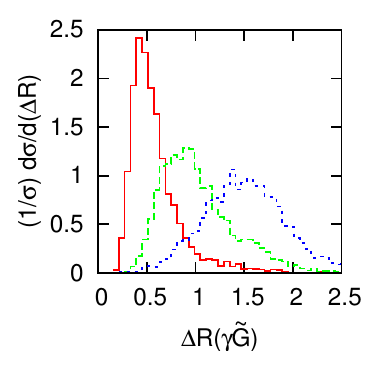}
\includegraphics[width=0.49\linewidth]{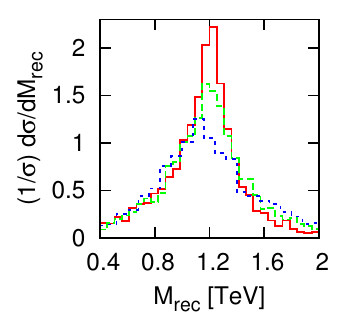}
\caption{
As Fig.~\ref{fig:1TeV}, holding the squark mass at $1.2$~TeV while varying the neutralino mass: $m_{\tilde{N}_1} = 100$,$\,200$,$\,400$~GeV are shown with red solid, green dashed, and blue dotted lines respectively.
The greater $m_{\tilde{N}_1}$, the less collinear its photon and gravitino daughters become, as shown by $\Delta R(\gamma \tilde{G})$ (averaged between the two $\gamma \tilde{G}$ pairs) in the left panel.
This worsens the mass reconstruction: the right panel shows one of the two masses found using one of the two jet-neutralino pairing criteria (all four quantities behave similarly -- see  Fig.~\ref{fig:1TeV}).
Events are generated with {\tt Herwig++}.
}
\label{fig:BoostedOrNot}
\end{figure}
\end{center}

{\it A Less Boosted Intermediate.}
Collinearity of $\chi$ and ${\rm vis}_2$ relies on their common mother particle being boosted; as it becomes less boosted they become less collinear.
We show this effect, and the decreasing sharpness of the mass reconstruction that results, in Fig.~\ref{fig:BoostedOrNot} for our previous gauge-mediation example.
$m_{\tilde{N}_1}$ is increased from $100$ to $400$~GeV for constant $m_{\tilde{q}} = 1.2$~TeV.
If $\tilde{N}_1$ is made heavier still, e.g. $m_{\tilde{N}_1}/m_{\tilde{q}}\rightarrow1$, the increasingly lethargic neutralino gives a less collimated photon-gravitino pair; indeed the two are increasingly back-to-back, and most events fail to meet the requirement that $\slashed{\mathbf{p}}_T$ be in between the two photons.

{\it More Decays Of The Intermediate.}
If ${\rm vis}_2$ is several particles instead of the single photon $\gamma$ we considered, e.g. a lepton pair from a boosted $\tilde{N}_i\rightarrow l^{\pm}l^{\mp}\tilde{N}_1$ decay, by construction they will be collimated and the sum of their four-momenta can be used in place of $p_{\gamma}^{\mu}$ in the analysis.

\begin{center}
\begin{figure}[!ht] 
\centering
\includegraphics[width=0.49\linewidth]{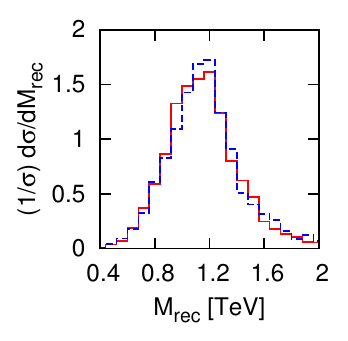}
\includegraphics[width=0.49\linewidth]{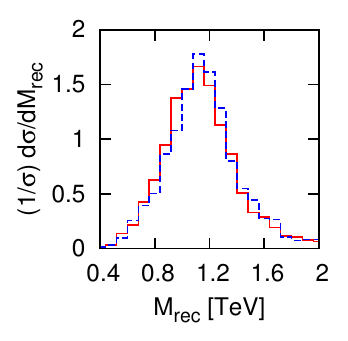}
\caption{
The gluino mass ($m_{\tilde{g}}= 1.2~\text{TeV}$) in the process $pp\rightarrow2\tilde{g}\rightarrow 4q+2(\tilde{N}_1)\rightarrow 4q+2(\tilde{G}+\gamma)$ reconstructed with missing energy decomposition and neutralino-jet pairing criterion $\beta$ as described in the text.
The masses of the lightest neutralino and gravitino are $m_{\tilde{N}_1}=100~\text{GeV},\:m_{\tilde{G}}=1~\text{eV}$ (the masses of other particles are set at $2$~TeV); the centre of mass energy is $8$~TeV.
Panels on the left (right) show the mass of the gluino calculated from the leading (sub-leading) photon in each event.
The blue dashed line shows events generated by {\tt MadGraph} and {\tt PYTHIA}, with fast detector simulation performed by {\tt PGS}; the red solid line shows events generated by {\tt Herwig++}.
}
\label{fig:gluino}
\end{figure}
\end{center}

{\it More Decays Before The Intermediate.}
If the directly pair-produced particles decay to a boosted intermediate and two visible particles rather than one -- via two on-shell steps or a three-body decay -- then each ${\rm vis}_1$ in Fig.~\ref{fig:Diagram} is replaced by two particles which are not collinear.
Criterion $\alpha$ is then not applicable but criterion $\beta$ is, albeit with greater combinatorial ambiguity from the need to pair each reconstructed neutralino with two other visible objects.
In this scenario the boosted intermediate is also less boosted from sharing its energy with more particles, making its semi-invisible decay less collimated.
Despite these difficulties the method is reasonably successful: Fig.~\ref{fig:gluino} shows events for a simplified model with pair-produced gluinos of mass $1.2$~TeV decaying to $q\bar{q}\tilde{N}_1$ ($q$ now denoting a quark of any of the three generations) with the $100$~GeV neutralino decaying to $\gamma \tilde{G}$.
Neutralino-jet pairing is performed with criterion $\beta$ generalised in the obvious way to include four jets rather than two.

{\it Other Combinatoric Complications.}
If ${\rm vis}_1 = {\rm vis}_2$, e.g. if in our former example photons were replaced by jets or jets by photons (but not both of these at once), then there would be a combinatoric ambiguity not just in pairing the reconstructed boosted intermediate with the correct ${\rm vis}_1$ but also in which two particles define the initial $\slashed{\mathbf{p}}_T$ decomposition directions.
The requirement that $\slashed{\mathbf{p}}_T$ be in between the two visible particles onto which it is decomposed eliminates some of the possible decomposition configurations; for the rest, criterion $\beta$ can be generalised to be an optimisation over decomposition configurations as well as pairing possibilities.

{\it More Than Two Invisible Particles.}
With a third $\chi$ in the final state which is expected to be \mbox{(anti-)}parallel to one of the first two, our ansatz for the topology still contains only two invisible directions and we can uniquely decompose the observed MET.
If the two invisible particles that are \mbox{(anti-)}parallel have come from the decay of the same particle, we only need to know the sum of their momenta and so we can reconstruct the mass as before.
However if they have come from the decay of two different particles, then we need their individual momenta for mass reconstruction; knowing only their sum, the masses we wish to calculate are underconstrained by one parameter.
Another possibility is
%In this case there may be non-unique MET decomposition.
$3\chi$ in the signal final state with three different expected directions: there are then three vectors in the transverse plane into which $\slashed{\mathbf{p}}_T$ can be decomposed, with any two of the three giving a unique decomposition.
There are three ways to choose two vectors from the three.
We may have the $\slashed{\mathbf{p}}_T$ in between the two vectors in $0$, $1$ or $2$ of the three ways (neglecting the possibility of exact collinearity between $\slashed{\mathbf{p}}_T$ and one of the vectors).
If $0$, we veto.
If $1$, there is a unique decomposition.
If $2$, $\slashed{\mathbf{p}}_T$ can be expressed as some amount of one of the decompositions plus some amount of the other, with the two coefficients constrained to sum to unity: the masses we wish to calculate are under-constrained by one parameter.
One response, not physically motivated, would be to veto.
Which of these three cases ($0$, $1$ or $2$ of the possible decompositions being acceptable) we have will vary on an event by event basis.

\section{Conclusion}

Partially invisible decays occur in many extensions of the Standard Model, being more or less omnipresent in models with dark matter candidates.
However when missing energy arises from two invisible particles, reconstruction of mass peaks -- and hence discovery -- is much more difficult, even though Standard Model backgrounds may not be dominating the signal.
We consider the case where there are two preferred directions for the two invisible particles, which may arise due to spin correlations, back-to-back decays or boosted decays.
The observed missing energy vector can then be decomposed into components along each of those directions; these components approximately describe the two invisible particles and may allow reconstruction of the mass of their mother particles.

Pairs of collimated semi-invisible decays together with jets occur in supersymmetry when heavy squarks or gluinos decay to light MSSM-like neutralinos, which decay in turn to gravitinos, pseudo-goldstini, singlinos or photini.
We considered as a concrete example the gauge-mediation style decay of $1.2$~TeV squarks or gluinos to jets, photons and gravitinos.
When the mass of the intermediate neutralino is $100$~GeV, the initial mass is reconstructed to $10\%$ accuracy for roughly $\tfrac{1}{3}$ of events passing the basic cuts.
Multiplying by the Prospino~\cite{Beenakker:1996ed,Beenakker:1996ch} production cross-section and the acceptance -- $20\,{\rm fb}\times0.5$ for squarks, $2.5\,{\rm fb}\times0.3$ for gluinos -- one would expect $\mathcal{O}(100)$ events for squarks, $\mathcal{O}(10)$ for gluinos, in $30\,{\rm fb}^{-1}$ at $8$~TeV.
(These numbers of events inside the peak of course depend on the masses -- when the mass of the neutralino approaches the mass of the squark/gluino, it is no longer boosted and a peak will not be reconstructed with this method.)

We have discussed how the method is applicable to final states with particles different from those in the example, and outlined the limitations as it is applied to more general scenarios.
The level of reconstruction is extremely encouraging and we hope that our results are an incentive for the experimental collaborations to investigate the feasibility of implementing such analyses.

\section{Acknowledgements}
CW thanks Olivier Mattelaer for helpful communications concerning {\tt MadGraph}, Ulrich Ellwanger for arranging extended hospitality at LPT Orsay, and the organisers of the Carg\`{e}se International School 2012 for the stimulating environment where this idea took shape.
This work was supported by the STFC.

\bibliographystyle{h-physrev3}
\bibliography{bib}

\begin{thebibliography}{10}

\bibitem{Barger:1987re}
V.~D. Barger, T.~Han, and J.~Ohnemus,
\newblock Phys.Rev. {\bf D37}, 1174 (1988).
%%CITATION = PHRVA,D37,1174;%%

\bibitem{Lester:1999tx}
C.~Lester and D.~Summers,
\newblock Phys.Lett. {\bf B463}, 99 (1999), hep-ph/9906349.
%%CITATION = HEP-PH/9906349;%%

\bibitem{Rogan:2010kb}
C.~Rogan,
\newblock (2010), 1006.2727.
%%CITATION = ARXIV:1006.2727;%%

\bibitem{Allanach:2000kt}
B.~Allanach, C.~Lester, M.~A. Parker, and B.~Webber,
\newblock JHEP {\bf 0009}, 004 (2000), hep-ph/0007009.
%%CITATION = HEP-PH/0007009;%%

\bibitem{Giudice:1998bp}
G.~Giudice and R.~Rattazzi,
\newblock Phys.Rept. {\bf 322}, 419 (1999), hep-ph/9801271.
%%CITATION = HEP-PH/9801271;%%

\bibitem{Kats:2011qh}
Y.~Kats, P.~Meade, M.~Reece, and D.~Shih,
\newblock JHEP {\bf 1202}, 115 (2012), 1110.6444.
%%CITATION = ARXIV:1110.6444;%%

\bibitem{Cheung:2010mc}
C.~Cheung, Y.~Nomura, and J.~Thaler,
\newblock JHEP {\bf 1003}, 073 (2010), 1002.1967.
%%CITATION = ARXIV:1002.1967;%%

\bibitem{Argurio:2011hs}
R.~Argurio, Z.~Komargodski, and A.~Mariotti,
\newblock Phys.Rev.Lett. {\bf 107}, 061601 (2011), 1102.2386.
%%CITATION = ARXIV:1102.2386;%%

\bibitem{Argurio:2011gu}
R.~Argurio {\em et~al.},
\newblock JHEP {\bf 1206}, 096 (2012), 1112.5058.
%%CITATION = ARXIV:1112.5058;%%

\bibitem{Hall:2011aa}
L.~J. Hall, D.~Pinner, and J.~T. Ruderman,
\newblock JHEP {\bf 1204}, 131 (2012), 1112.2703.
%%CITATION = ARXIV:1112.2703;%%

\bibitem{Ellwanger:2009dp}
U.~Ellwanger, C.~Hugonie, and A.~M. Teixeira,
\newblock Phys.Rept. {\bf 496}, 1 (2010), 0910.1785.
%%CITATION = ARXIV:0910.1785;%%

\bibitem{Maniatis:2009re}
M.~Maniatis,
\newblock Int.J.Mod.Phys. {\bf A25}, 3505 (2010), 0906.0777.
%%CITATION = ARXIV:0906.0777;%%

\bibitem{Das:2012rr}
D.~Das, U.~Ellwanger, and A.~M. Teixeira,
\newblock JHEP {\bf 1204}, 067 (2012), 1202.5244.
%%CITATION = ARXIV:1202.5244;%%

\bibitem{Baryakhtar:2012rz}
M.~Baryakhtar, N.~Craig, and K.~Van~Tilburg,
\newblock JHEP {\bf 1207}, 164 (2012), 1206.0751.
%%CITATION = ARXIV:1206.0751;%%

\bibitem{Macesanu:2002db}
C.~Macesanu, C.~McMullen, and S.~Nandi,
\newblock Phys.Rev. {\bf D66}, 015009 (2002), hep-ph/0201300.
%%CITATION = HEP-PH/0201300;%%

\bibitem{Cohen:2011aa}
T.~Cohen, A.~Hook, and B.~Wecht,
\newblock Phys.Rev. {\bf D85}, 115004 (2012), 1112.1699.
%%CITATION = ARXIV:1112.1699;%%

\bibitem{Plehn:1999xi}
T.~Plehn, D.~L. Rainwater, and D.~Zeppenfeld,
\newblock Phys.Rev. {\bf D61}, 093005 (2000), hep-ph/9911385.
%%CITATION = HEP-PH/9911385;%%

\bibitem{Englert:2012wf}
C.~Englert, M.~Spannowsky, and C.~Wymant,
\newblock Phys.Lett. {\bf B718}, 538 (2012), 1209.0494.
%%CITATION = ARXIV:1209.0494;%%

\bibitem{CMSsearch}
{CMS Collaboration},
\newblock Physics Analysis Summary CMS-PAS-SUS-12-018.

\bibitem{Allanach:2001kg}
B.~Allanach,
\newblock Comput.Phys.Commun. {\bf 143}, 305 (2002), hep-ph/0104145.
%%CITATION = HEP-PH/0104145;%%

\bibitem{Bahr:2008pv}
M.~Bahr {\em et~al.},
\newblock Eur.Phys.J. {\bf C58}, 639 (2008), 0803.0883.
%%CITATION = ARXIV:0803.0883;%%

\bibitem{Alwall:2011uj}
J.~Alwall, M.~Herquet, F.~Maltoni, O.~Mattelaer, and T.~Stelzer,
\newblock JHEP {\bf 1106}, 128 (2011), 1106.0522.
%%CITATION = ARXIV:1106.0522;%%

\bibitem{Sjostrand:2006za}
T.~Sjostrand, S.~Mrenna, and P.~Z. Skands,
\newblock JHEP {\bf 0605}, 026 (2006), hep-ph/0603175.
%%CITATION = HEP-PH/0603175;%%

\bibitem{PGS}
{J. Conway {\it et al}},
\newblock
  \url{http://www.physics.ucdavis.edu/~conway/research/software/pgs/pgs4-gener%
al.htm}.

\bibitem{Cacciari:2011ma}
M.~Cacciari, G.~P. Salam, and G.~Soyez,
\newblock Eur.Phys.J. {\bf C72}, 1896 (2012), 1111.6097.
%%CITATION = ARXIV:1111.6097;%%

\bibitem{Buckley:2010ar}
A.~Buckley {\em et~al.},
\newblock (2010), 1003.0694.
%%CITATION = ARXIV:1003.0694;%%

\bibitem{Me}
\url{http://www.ippp.dur.ac.uk/~hndv85/}.

\bibitem{Cacciari:2008gp}
M.~Cacciari, G.~P. Salam, and G.~Soyez,
\newblock JHEP {\bf 0804}, 063 (2008), 0802.1189.
%%CITATION = ARXIV:0802.1189;%%

\bibitem{Lester:2011nj}
C.~G. Lester,
\newblock JHEP {\bf 1105}, 076 (2011), 1103.5682.
%%CITATION = ARXIV:1103.5682;%%

\bibitem{Kribs:2009yh}
G.~D. Kribs, A.~Martin, T.~S. Roy, and M.~Spannowsky,
\newblock Phys.Rev. {\bf D81}, 111501 (2010), 0912.4731.
%%CITATION = ARXIV:0912.4731;%%

\bibitem{Alwall:2007fs}
J.~Alwall {\em et~al.},
\newblock Eur.Phys.J. {\bf C53}, 473 (2008), 0706.2569.
%%CITATION = ARXIV:0706.2569;%%

\bibitem{Beenakker:1996ed}
W.~Beenakker, R.~Hopker, and M.~Spira,
\newblock (1996), hep-ph/9611232.
%%CITATION = HEP-PH/9611232;%%

\bibitem{Beenakker:1996ch}
W.~Beenakker, R.~Hopker, M.~Spira, and P.~Zerwas,
\newblock Nucl.Phys. {\bf B492}, 51 (1997), hep-ph/9610490.
%%CITATION = HEP-PH/9610490;%%

\end{thebibliography}

\end{document}